\documentclass[12pt]{article}
\usepackage{url}
\usepackage{amsmath}
\usepackage[top=1.0in,bottom=1.0in,left=1.25in,right=1.25in]{geometry}

\begin{document}
\bibliographystyle{acm}

\date{July 6, 2007}
\author{John Mount\thanks{http://www.mzlabs.com/},
Nina Zumel\thanks{http://www.quimba.com/}}
\title{
Comparing Apples and Oranges:
Two Examples of the Limits of Statistical Inference,
With an Application to Google Advertising Markets}
\maketitle

\section{Overview}

\footnotetext{
This work is licensed under the Creative Commons Attribution-Share
Alike 3.0 United States License. To view a copy of this license, visit
http://creativecommons.org/licenses/by-sa/3.0/us/ or send a letter to
Creative Commons, 171 Second Street, Suite 300, San Francisco,
California, 94105, USA.
}

Bad experimental situations are often a source of great statistical
puzzles.  We are going to describe an example of this sort of situation using what
one author observed while watching a few different companies using the
Google AdSense and AdWords products.

The points we argue will be obvious to statisticians -- in fact, they are
actually elementary exercises.  We will show that the measurements
allowed in the Google AdSense markets are insufficient to allow
accurate tracking of a large number of different revenue sources.

Our goal is to explain a well known limit on inference to a larger
non-specialist audience.  
This is a bit of a challenge as most
mathematical papers can only be read by people who could have written
the paper themselves.  
By ``non-specialist audience'' we mean analytically minded
people that may not have seen this sort of math before, or those who
have seen the theory but are interested in seeing a complete application.
We will include in this writeup the notes,
intents, side-thoughts and calculations that mathematicians produce to
understand even their own work but, as Gian-Carlo Rota wrote, we are
compelled to delete for fear our presentation and understanding won't
appear as deep as everyone else's.\cite{rota:1997a} 

The counter-intuitive points that we wish to emphasize are:

\begin{itemize}
\item The difficulty of estimating the variance of individuals from a small number of aggregated measurements.
\item The difficulty of estimating the averages of many groups from a small number of aggregated measurements.
\end{itemize}

These points will be motivated as they apply in the Google markets
and we will try to examine their consequences in a simplified setting.

\pagebreak

\tableofcontents

\pagebreak

\section{The Google Markets}

\subsection{Introduction}

Google both buys and sells a large number of textual advertisements through
programs called Google AdSense and Google AdWords.\cite{goog1}  What is
actually purchased and sold is ``clicks.''  Web sites that agree to
display Google AdSense are paid when users click on these ads, and
advertisers who place advertisements into Google AdWords pay
Google when their advertisements are clicked on.  The key item in
these markets is the ``search term'' that the advertiser chooses to
bid on advertising clicks for. ``Search terms'' are short phrases for which
an advertiser is willing to pay, in order to get a visit from a web surfer
who has performed
a search on that phrase. For instance a company
like Panasonic might consider clicks on the search term  ``rugged laptop''
(and the  attention of the underlying web surfer) 
to be worth \$2 to them.

Because Google both buys and sells advertisements they are essentially
making a market.  There are some unique aspects to this market in that
it is not the advertisements or even page-views that are being traded,
but clicks.  Both Google and its affiliates serve
the advertisements for free and then exchange payment only when a
web surfer clicks on an advertisement. 
A website can ``resell'' advertisements by simultaneously placing
ads through AdWords, and serving ads through AdSense. When a
user clicks into the website via an advertisement, this costs the
web site money; if, however, the user is then shown a number of other
advertisements, he or she may then click out on one of
them of their own free will, recouping money or perhaps even making a profit for the site.  
There is significant uncertainty
in attempting resale and arbitrage in these advertisement markets, as the
user who must be behind all the clicks can just ``evaporate'' 
during an attempted resale.  Direct reselling of clicks
(such as redirecting a web surfer from one advertisement to another)
would require a method called ``automatic redirection''
to move the surfer from 
one advertisement to a replacement advertisement.  Automatic
redirection is not allowed by Google's terms of service.

An interesting issue is that each click on a given search term is
a unique event with a unique cost.  One click for ``rugged laptop''
may cost \$1 and another may cost \$0.50.  
The differing costs are
determined by the advertiser's bid, available placements for the key
phrase, what other advertisers are bidding in the market, how many
web surfers are available, and Google's sorting of bids.
The sorting of bids by Google depends on the rank of advertiser's bid
times an adjustment factor managed by Google.
The hopeful assumption is
that all of the potential viewers and clickers for the same search term
are essentially exchangeable in that they all have a similar (unknown)
cost and similar probabilities of later actions, such as
buying something from a web site.
The concept of exchangeability is
what allows information collected on one set of unique events to inform
predictions about new unique events (drawn from the same exchangeable
population).  

Whatever the details are, these large advertisement markets have given
Google an income of \$12 billion, \$3.5 billion in profit and 70\%
year to year growth in 2006.\cite{googval}  This scale of profit is
due in part to the dominant position of Google in forming markets for
on-line advertising.  

The reasons for Google's market domination are various and include the
superior quality of the Google matching and bidding service, missteps by
competitors and the network effects found in a good market -- the
situation whereby sellers attract buyers and buyers attract sellers.
The cost of switching markets (implementation, information handling and staffing
multiple relationships) are also significant factors. 

In our opinion, Google's profit margins are also helped by
the limits on information available to most of the other market 
participants. In the next section, we will discuss some of the information limits or
barriers to transparency in the Google market.

\subsection{Information Limits}

Google deals are typically set up as revenue sharing arrangements in
which Google agrees to pay a negotiated portion of the revenues
received by Google to the AdSense hosting web site.  As noted above,
advertisement click-through values vary from as little as \$0.05 to
over \$40.0 per click.  It is obvious that web site operators who
receive a commission to serve advertisements on behalf of the Google
AdSense program need detailed information about which advertisements
are paying at what rate.  This is necessary both to verify that Google
is sharing the correct amount on valuable advertisements and to adjust
and optimize the web site hosting the advertisements.

However, Google does not provide AdSense participants with a complete
breakdown of revenues paid.
There are a number of possible legitimate reasons for 
this. First, there is a concern that allowing web sites complete
detailed reconciliation data would allow them to over-optimize or
perform so-called ``keyword arbitrage'' where sites buy precisely the
keywords they can profitably serve advertisements on instead of buying
keywords for which the site actually has useful information or services.
In addition, the quantity of data is very large, so there are some technical
challenges in providing a detailed timely reconciliation.  There can
also be reasons favorable to Google.

\subsection{Channel Identifiers}

Google's current solution to the conflicting informational needs defines the
nature of the market and is in itself quite interesting.  Google
allows the AdSense customer a number of measurements called
``channels.''  The channels come with identifiers and the AdSense
customer is allowed to attach a number of identifiers to every advertisement
clicked-out on.  Google in turn reports not the detailed revenue for
every click-out but instead just the sum of revenue received on
clicks-out containing each channel identifier.

For example: if a web site operator wanted to know the revenue from a
particular search term (say ``head cold'') they could attach a
single channel identifier to all click-outs associated with ``head
cold'' and to no other search term. Under this scheme, Google would then
be reporting the revenue for the search term as a channel summary.
This simple scheme uses up an entire
channel-id for a single search term.  This would not be a problem
except that an AdSense partner is typically limited (by Google) to a
few hundred channel identifiers and is often attempting to track tens
of thousands of search terms (and other conditions such as traffic source and
time of day).  It is obvious to any statistician that
these limited number of channels are not sufficient to eliminate 
many degrees of uncertainty in the revenue attribution problem.

Google does allow each click-out to have multiple channel identifiers
attached to it.  At first this seems promising -- for instance one can
easily come up with schemes where 30 channel ids would be sufficient to give
over a billion unique search terms each a unique {\em pattern} of
channel identifiers.  However, Google does not report revenue for each
pattern of channel identifiers; in this case they would only report
the total for each of the 30 channels.  Each channel total would be the sum of
all revenue given for all clicks-out that included the given
channel-id.  Under this scheme we would have a lot of double counting
in that any click-out with multiple channel identifiers attached is
necessarily simultaneously contributing to multiple totals.  Anyone
familiar with statistics or linear algebra will quickly recognize that
30 channels can really only reliably measure about 30 facts about an
ad campaign.  {\em There is provably no super clever scheme capable of
decoding these confounded measurements into a larger number of
reliable outcomes}. 

Let us go back to the points that we promised
to discuss at the beginning of this paper:

\begin{itemize}
\item The difficulty of estimating the variance of individuals from a small number of aggregated measurements.

In terms of Google AdSense, this means that we can tell the average (mean) value of a click in a given channel, 
but we cannot tell how widely the click values in the channel vary from this average value.

\item The difficulty of estimating the averages of many groups from a small number of aggregated measurements.

This means that if we assign multiple search terms into each of our available channels, we cannot separate out the values of
each individual search term using only the aggregate channel measurements.
\end{itemize}

It is an interesting exercise to touch on the theory of why these facts are true.

\section{The Statistics}

One thing the last section should have made obvious is that even
describing the problem is detailed and tedious.  It may be better to
work in analogy to avoid real-world details and non-essential complications.  
Let's replace advertisement clicks-out
with fruit, and channels with weighings of baskets.

Suppose we are dealing with apples and our business
depends on knowing the typical weight of each fruit.  We assume that
all apples are exchangeable: they may each have a different weight (and
value) but they all are coming from a single source.  We further
assume that we have a limited number of times that we are allowed to place
our apples into a basket and weigh them on a scale.

\subsection{The Variance is Not Measurable}

\subsubsection{The Mean\label{sec:themean}}

The first example, the happy one, is when we have a single basket
filled with many different items of one type of fruit.  For instance
suppose we had a single basket with 5 apples in it and we were told
the basket contents have a total weight of 1.3 pounds.  The fact that we were
given only a single measurement for the entire basket (instead of
being allowed to weigh each apple independently) does not interfere in
any way with accurately deducing that the average (or mean) of this
type of apple weighs a little more than 1/4 pound.   If we had $n$
apples in the basket, and we called the total weight of the contents of the
basket $T$, we could estimate the average or mean weight of
individual apples as being $T/n$.  If we use $a_w$ to denote the
(unknown universal)
average weight of individual apples we would denote our estimate of
this average as $\hat{a}_w$ and we have just said that our estimate
is $\hat{a}_w = T/n$.

However, we are missing the opportunity to learn at least
one important thing:
how much does the weight of these apples vary?  This could be an
important fact needed to run our business (apples below a given weight
may be unsellable, or other weight considerations may apply).  We may
need to know how inaccurate is it to use the mean or average weight of
the apples in place of individual weights.

If we were allowed 5 basket weighings we could put one apple in each
basket and directly see how much the typical variation in weight is
for the type of apples we have. Let's call this {\em Experiment-A}. 
Suppose in this case we find the 5
apples to weigh $0.25lb, 0.3lb, 0.27lb, 0.23lb, 0.25lb$ respectively.
This detailed set of measurements helps inform us on how this type of
apple varies in weight.  

One of the simplest methods to summarize
information about variation is a statistical notion called
``variance.''  Variance is defined as the expected squared distance of
an random individual from the population average.  Variance is written
as $E[(x - a_w)^2]$ where $x$ is a ``random variable'' 
denoting the weight of a single apple drawn
uniformly and independently at random (from the unknown larger population) 
and the $E[]$
notation denotes ``expectation.''  $E[(x - a_w)^2]$ is
the value that somebody who knew the value of $a_w$ would say is the
average value of $(x - a_w)^2$ over very many repetitions of drawing a
single apple and recording its individual weight as $x$.  For example
if all apples had the exact same weight the variance would be zero.

For the basket above, 
$E[(x - \hat{a}_w)^2]$ is calculated as:
\[
\frac{(0.25-0.26)^2 + (0.3-0.26)^2 + (0.27-0.26)^2 + (0.23-0.26)^2 + (0.25-0.26)^2}{5}
\]
(the $0.26$ itself the average of the 5 apples weights).
The interpretation is that for a similar apple
with unknown weight $x$ we would expect $(x-0.26)^2 \approx 4 * 0.00056$ or
for $x$ to not be too far outside the interval $0.212$ to $0.307$
(applying the common rule of thumb ``2 standard deviations''
which is 4 variances).  As we see all of the 
original 5 apples fell in this interval.

Now the 5 apple weights we know are not actually all the possible apples in
the world, they are merely the apples in our sample.  There are some
subtleties about using the variance found in a sample to estimate the
variance of the total population, but for this discussion we will use
the naive assumption that they are nearly the same.  If we use the
symbol $v_a$ to denote the (unknown) true variance of individual apple
weights (so $v_a = E[(x - a_w)^2]$) we can use it to express the fact
$\hat{a}_w$ is actually an excellent estimate of $a_w$.  

Specifically: if we were to repeat the experiment of taking a basket
of randomly selected apples ($n$ apples in the basket) over and over again, 
estimating the mean apple weight $\hat{a}_w$ each time, then 
$E[(\hat{a}_w - a_w)^2]$ -- the expected square error between
our estimate of the average apple weight and the true average apple weight -- will go to zero
as the sample-size $n$  is increased.  In fact, we can show
$E[(\hat{a}_w - a_w)^2] =  v_a/n$, which means that our estimate of the mean
gets more precise as $n$ is increased.  This fact that large samples are very
good estimates of unknown means is basic- but for completeness
we include its derivation in the appendix.

\subsubsection{Trying to Estimate the Variance}

We introduced the variance of individual apples (denoted by $v_a$) as
an unknown quantity that aided reasoning.
We know that even with only one measurement
of the total weight of all $n$ apples that $\hat{a}_w$ is an estimate
of the mean whose error goes to zero as the $n$ (the number of apples or
the sample size) gets large.

However, the variance of individual apples $v_a$ is so useful that we would
like to have an actual estimate ($\hat{v}_a$) of it.  
It would be very useful to know
if $v_a$ is near zero (all apples have nearly identical weight) or
if $v_a$ is large (apples vary wildly in weight).  
If we were allowed to weigh each apple as in Experiment-A (i.e. if we
had an unlimited number of basket weighings or channels), we could
estimate the variance by the calculations in the last section.
If we were allowed only one measurement
we would really have almost no information about the variance
as we have only seen one aggregated measurement- so we have no idea
how individual apple weights vary.  The next question is: can we create a
good estimate $\hat{v}_a$ when we are allowed only two measurements
but the sample size ($n$) is allowed to grow?

Lets consider {\em Experiment-B}:
If we have a total of $2 n$
apples ($n$ in each basket) and $T_1$ is the total weight of the first
basket and $T_2$ is the total weight of the second basket then 
some algebra would tell us that
$\hat{v}_a = \frac{(T_1-T_2)^2}{2 n}$ is an unbiased estimate of 
$v_a$ (the variance in weight of individual apples)\footnote{  
``Unbiased'' simply means that $E[\hat{v}_a - v_a] = 0$ which can also
be written as $E[\hat{v}_a] = v_a$.
This means our estimate of variance doesn't tend to be more over than
under (or more under than over).}.  

It turns out, however,  that $\hat{v}_a$ is actually a bad estimate of the
variance.  That is, the expected distance of $\hat{v}_a$ from the unknown 
true value of the variance $v_a$ (written $E[(v_a - \hat{v}_a)^2]$) 
does not shrink beyond a certain bound as the number of apples in each 
basket ($n$) is increased.  This ``variance of variance estimate'' result
is in stark contrast to the nice behavior we just saw in estimating the
average $a_w$.  With some additional assumptions and algebra (not shown here) we can show that 
for our estimate $\hat{v}_a = \frac{(T_1 - T_2)^2}{2 n}$ we have
$\lim_{n \rightarrow \infty} E[(\hat{v}_a - v_a)^2] = 2 v_a^2$.
There is a general 
reason this is happening, and we will discuss 
this in the next section.


\subsubsection{Cramer-Rao: Why we can not estimate the variance of individual Apples}

Of course showing one particular calculation fails is not the same as
showing that the variance of individual apples can not be estimated
from the two total weighings $T_1$ and $T_2$.  There could be other,
better, estimates\footnote{As an aside, some of the value in proposing a specific estimate
(because the theory says there is no good one) is that it allows
one to investigate the failure of the estimate without resorting
to the larger theory.  
For example in this day of friendly computer languages and ubiquitous computers
one can easily empirically confirm (by setting up a simulation 
experiment as suggested by Metropolis and Ulam\cite{Metropolis:1949:MCM}).
One can check that our estimate is unbiased (by averaging
many applications of it) and that it is not good (by observing the
substantial error on each individual application even when $n$ is
enormous).  There is no rule that one should not get an empirical
feel (or even an empirical confirmation) of a mathematical statement
(presentation of math is subject to errors) and in this day there 
are likely many more readers who could quickly confirm or disprove
the claims of this section by simulation than there are readers who
would be inclined to check many lines of tedious algebra for 
a subtle error.
}.

There is a well known statistical law that states no
unbiased estimator works well in this situation.
The law is called the Cramer-Rao inequality.\cite{cove_thom_91}
The Cramer-Rao inequality
is a tool for identifying situations where {\em all} unbiased estimators
have large variance.  The Cramer-Rao inequality is typically a calculation so we
will add a few more (not necessarily realistic) assumptions to ease calculation.
We assume apple weights are distributed normally with mean $a_w$ and variance
$v_a$.\footnote{``Normal'' is a statistical term for the distribution
associated with the Bell curve.  Many quantities in nature have a nearly normal
distribution.}

There is a quantity depending only on the experimental set up 
that reads off how difficult estimation is.
By ``depending only on the experimental set up'' we mean that 
the quantity does not depend on any specific outcomes of $T_1$, $T_2$ 
and does not depend on
any specific estimation procedure or formula.  This quantity is
called ``Fisher Information'' and is denoted as $J(v_a)$.

The Cramer-Rao inequality\cite{cove_thom_91}
says for any unbiased estimator $\hat{v}$,
the variance of $\hat{v}$ is at least $1/J(v_a)$.  Written in
formulas the conclusion of the Cramer-Rao inequality is:
\[
E[(v_a - \hat{v})^2] \ge 1/J(v_a)
.\]

Since we have
now assumed a model for the weight distribution of apples,
we can derive (see appendix) the following:
\[
J(v_a) = \frac{2}{v_a^2}
.\]

Applying the Cramer-Rao inequality lets us immediately say:
\[
E[(v_a - \hat{v})^2] \ge \frac{v_a^2}{2}
.\]

This means that there is no  unbiased estimation procedure for which can we expect the
squared-error to shrink below $\frac{v_a^2}{2}$ even as the
number of items in each basket ($n$) is increased.  So not only does our
proposed variance estimate fail to have the (expected) good behavior 
we saw when estimating the mean, but in fact no unbiased estimating scheme will work.  In general we can show that the quality of the variance estimate
is essentially a function of the number of measurements we are 
allowed\footnote{And perhaps surprisingly not a function of the sample size.}
- so any scheme using a constant number of measurements will fail.

\subsection{Trying to Undo a Mixture\label{sec:mixture}}

Suppose we are willing to give up on estimating the variance (a dangerous
concession).  We are still blinded by the limited number of channels
if we attempt to estimate more than one individual mean.

In our analogy let's introduce a second fruit (oranges) to the problem.
Call an assignment of fruit to baskets a ``channel design.''  For
example if we were allowed two basket measurements and wanted to know
the mean weight of apples and the mean weight of oranges we could assign
all apples to one basket and all oranges to the other.  This ``design''
would give us very good estimates of both the mean weight of apples and
the mean weight of oranges.

Let's consider a simple situation where due to the limited number
of channels we are attempting to measure something that was not considered
in the original channel design.  This is very likely because the number
of simultaneous independent measurements is limited to the number of
channels and it is very likely that one will have important questions that
were not in any given experimental design.   
For example (going back to AdSense),
suppose we had
26 channels and we used them all to group our search phrases by
first letter of the English alphabet and we later wanted to break down 
older data by length of phrase.\footnote{These examples are deliberately 
trivial.}   We would consider ourselves lucky if the
first-letter design was even as good as random assignment of channel ids
in measuring the effect of search term length.

To work this example we continue to ignore most of the 
details and suppose we really are
trying to estimate the mean weight of apples and the mean weight of oranges
at the same time.  Due to the kind of bad luck described above we have data
from an experiment that was not designed for this purpose.  
Let's try the so-called easy case where we have a 
random experiment.  For {\em Experiment-C }
let's suppose we have two baskets of fruit and each basket
was filled with $n$-items of fruit by repeating the process of flipping a 
fair coin and placing an apple if the coin came up heads and an orange if
the coin came up tails.  This admittedly silly process is simulating the
situation where we are forced to use measurements that potentially could solve
our problem- but were not designed to solve it.\footnote{
This is one of the nasty differences between prospective studies
where the experimental layout is tailored to expose the quantities of
interest and retrospective studies where we hope to infer new quantities
from experiments that have relevant (but not specifically organized) data.}  
We can measure the total
weight of the contents of each basket.  So the information at our 
disposal this time is $a_1,o_1,T_1$ (the number of apples in the first basket,
the number of oranges in the first basket and the total weight of the 
first basket) and $a_2,o_2,T_2$ (the number of apples in the second basket,
the number of oranges in the second basket and the total weight of the 
second basket).  What we want to estimate are $a_w$ and $o_w$  the unknown 
mean weights of the types of apples and types of oranges we are dealing with.

To simplify things a bit let's treat the number of apples and oranges in each basket,
$a_1,o_1,a_2,o_2$, as known constants set at
``typical values'' that we would expect from the coin flipping procedure.
It turns out the following values of $a_1,o_1,a_2,o_2$ are typical:

\begin{eqnarray*}
a_1 & = & n/2 + \sqrt{n} \\
o_1 & = & n/2 - \sqrt{n} \\
a_2 & = & n/2 - \sqrt{n} \\
o_2 & = & n/2 + \sqrt{n}.
\end{eqnarray*}

We call these values typical because in any experiment where the
distribution of $n$ items in a collection is chosen by fair coin flips
we expect to see a nearly even distribution (due to the fairness of the coin)
but not too even (due to the randomness).  In fact we really do expect
any one of these 
values to be at least $\sqrt{n}/2$ away from $n/2$ most of the time
and closer than $2 \sqrt{n}$ most of the time.
So these are typical values, good but not too good.

We illustrate how to produce an unbiased (though in the end unfortunately
unusable) estimate for $a_w$ and $o_w$.  The general theory says the
estimate will be unreliable- but there is some value in seeing how 
an estimate is formed and having a specific estimate to experiment with.
The fact that we know the count of each fruit in each basket, and 
each basket's weight, 
gives us a simultaneous system of
equations:

\begin{eqnarray*}
E_{a_1,o_1,a_1,o_2}[T_1] & = & a_1 a_w + o_1 o_w \\
E_{a_1,o_1,a_2,o_2}[T_2] & = & a_2 a_w + o_2 o_w
\end{eqnarray*}

$E_{a_1,o_1,a_1,o_2}[T_1]$ represents the average value of $T_1$ over imagined
repeated experiments where $a_1$ apples and $o_1$ oranges are placed
in a basket and weighed (similarly for $E_{a_1,o_1,a_2,o_2}[T_2]$).  
The subscripts are indicating we are only considering experiments
where the number of apples and oranges are known to be 
exactly $a_1,o_1,a_1,o_2$.
We do not actually
know $E_{a_1,o_1,a_1,o_2}[T_1]$ and $E_{a_1,o_1,a_2,o_2}[T_2]$ 
but we can use the specific basket total
weighs $T_1,T_2$ we saw in our single experiment 
as stand-ins.  
In other words, $T_1$ may not equal $E_{a_1,o_1,a_1,o_2}[T_1]$ but $T_1$
is an unbiased estimator of $E_{a_1,o_1,a_1,o_2}[T_1]$ (this is 
a variation on the old ``typical family with 2.5
children'' joke).  So we rewrite the previous system as estimates:

\begin{eqnarray*}
T_1 & \approx & a_1 a_w + o_1 o_w \\
T_2 & \approx & a_2 a_w + o_2 o_w .
\end{eqnarray*}

We can the rewrite this system into a ``solved form'':

\begin{eqnarray*}
a_w & \approx & \frac{o_2 T_1 - o_1 T_2}{a_1 o_2 - a_2 o_1} \\
o_w & \approx & \frac{-a_2 T_1 + a_1 T_2}{a_1 o_2 - a_2 o_1} .
\end{eqnarray*}

And this gives us the tempting estimates $\hat{a}_w$ and $\hat{o}_w$

\begin{eqnarray*}
\hat{a}_w & = & \frac{o_2 T_1 - o_1 T_2}{a_1 o_2 - a_2 o_1} \\
\hat{o}_w & = & \frac{-a_2 T_1 + a_1 T_2}{a_1 o_2 - a_2 o_1} .
\end{eqnarray*}

$\hat{a}_w$ and $\hat{o}_w$ are
indeed unbiased estimates of $a_w$ and $o_w$.

The problem is: even though these are unbiased 
estimates- they are not good estimates.  With some calculation 
one can show that as $n$ (the number of pieces of fruit in each basket)
increases that $E_{a_1,o_1,a_2,o_2}[(\hat{a}_w - a_w)^2]$ 
and $E_{a_1,o_1,a_2,o_2}[(\hat{o}_w - o_w)^2]$ do not approach zero.  Our estimates
have a certain built-in error bound that does not shrink even as the
sample size is increased.

\subsubsection{Cramer-Rao: Why we can't separate Apples from Oranges}

What is making estimation difficult has been the same in all
experiments: most of what we want to measure is being obscured.  As
we mentioned earlier, in
a typical case all of $a_1,o_1,a_2,o_2$ will be relatively near a common 
value.  Any estimation procedure is going to depend on separations 
among these values, which are unfortunately not that big.  This is what
makes estimation difficult.

Let us assume apple weights are distributed normally with mean $a_w$ and variance
$v_a = v$ and orange weights are distributed normally with mean $o_w$ and 
variance $v_o = v$.

Since we have
now assumed a model for the weight distribution of apples and oranges
we can derive (calculating as shown in \cite{cove_thom_91})
the following:
\[
J(a_w,o_w) = 
\frac{1}{n v}
\begin{bmatrix}
a_1^2 + a_2^2 & a_1 o_1 + a_2 o_2 \\
a_1 o_1 + a_2 o_2 & o_1^2 + o_2^2 \\
\end{bmatrix}
.\]

What we are really interested in is the inverse of $J(a_w,o_w)$, which 
(for or typical values of $a_1,o_1,a_2,o_2$) 
is:
\[
J^{-1}(a_w,o_w) = 
\frac{v}{8}
\begin{bmatrix}
1 + 4/n & -1 + 4/n \\
-1 + 4/n & 1 + 4/n
\end{bmatrix}
.\]

The theory says that 
the diagonal entries of this matrix are essentially lower bounds on 
the squared error 
in the estimates of the apple and orange weights, respectively. 
The off-diagonal terms
describe how an error in the estimate of the mean apple weight affects 
the estimate of the mean orange weight, and vice-versa. 
So what we would like is for 
all the entries of $J^{-1}(a_w,o_w)$ to approach zero as $n$ increases. 
In our case, however, the entries of $J^{-1}(a_w,o_w)$ all tend to the
constant $\frac{v}{8}$ as $n$ grows, meaning that the errors in the estimates
are also bounded away from zero and stop improving as the sample size 
increases.

The above discussion assumes that the distribution of apples and oranges in each
basket is the same (in this case, random and uniform). If there is some constructive bias 
in the process forming
$a_1,o_1,a_2,o_2$, such as apples being a bit
more likely in the first basket and oranges a bit more likely 
in the second basket, then the demonstrated estimate is good
(with error decreasing
as $n$ grows) and is actually useful.  But 
the degree of utility of the estimate depends on how much useful bias we
have- if there is not much useful bias then the errors shrink very slowly
and we need a lot more data than one would first 
expect to get a good measurement.
Finally, we would like to remind the reader that it is impossible for
a channel design with a limited number of channels to simultaneously have
an independent large useful bias on very many measurements.

As an example of the application of useful bias
suppose that our coin has probability $p$ of coming up heads, and that
the first basket is filled by placing an apple every time the coin is heads, and an orange
every time the coin is tails. The second basket is filled the opposite way -- apple for tails,
orange for heads.  Again, let's treat the number of apples and oranges in each basket,
$a_1,o_1,a_2,o_2$, as known constants set at
``typical values'' that we would expect from the coin flipping procedure.

\begin{eqnarray*}
a_1 & = & np \\
o_1 & = & n(1-p) \\
a_2 & = & n(1-p)  \\
o_2 & = & np
\end{eqnarray*}
(as long as $p \neq \frac{1}{2}$ the $\sqrt{n}$ terms are dominated by
the bias and can be ignored).

If $p=1$ -- the coin always comes up heads -- then the first basket is only apples, and
the second basket is only oranges, and obviously, we can find good estimates of $a_w$ and
$o_w$, by the arguments in Section \ref{sec:themean}.  If $p=\frac{1}{2}$, then we are in the situation that
we already discussed, with approximately equal numbers of apples and oranges in each basket.
But suppose $p$ were some other value besides $1$ or $\frac{1}{2}$, say, $p=\frac{1}{4}$. In that case,
the first basket would be primarily oranges, and the second one primarily apples, and we can show that

\[
J^{-1}(a_w,o_w) = 
\frac{v}{2n}
\begin{bmatrix}
5 & -3 \\
-3  & 5
\end{bmatrix}
,\]
and all of the entries of $J^{-1}(a_w,o_w)$ do go to zero as $n$ gets larger. This can be shown to be true in 
general, for any $p\ne\frac{1}{2}$.  This means the Cramer-Rao bound
does not prevent estimation.
Another calculation (not shown here) confirms that our proposed estimate
does indeed have shrinking error (as $n$ increases).

\section{Other Solution Methods}

We did not discuss solution methods that involve more data,
such as repeated experiments, or significantly deeper knowledge, such as
factor models.  What we discussed were the limits of
the basic modeling step, which itself would be a component of the 
more sophisticated solutions. 
Here however, we will briefly touch on other procedures that
could be used to try to improve the situation discussed above.

{\em Repeated measurements} could be implemented by taking data over
many days, reassigning the channel identifiers so that each search term participates in
different combinations of channel identifiers over the course of the measurements.
Essentially, this is setting up a much larger system of simultaneous equations, from which
a larger number of variables can be estimated. 
There are mathematical procedures for this sort of iterative
estimation (such as the famous Kalman filter), but the 
number of quantities a web site would wish to estimate 
is so much larger than the number of measurements available that the 
procedure will require many reconciliation rounds to converge. In addition, 
this model assumes that the values of the variables being measured do not 
change over time (or change very slowly). This is not an assumption that is 
necessarily true in the AdWords domain, due to seasonality and other effects.

A {\em factor model} is a model where one has researched a small number of
causes or factors that explain the expected value of search phrases in 
a very simple manner.
For example it would be nice if the value of a search phrase were
the sum of a value determined by the first letter plus an independent
value determined by the second letter.
In such a case we would only need $2*26 = 52$ 
channels (to track the factors) and we would then be able to apply 
our model to many different search phrases.  Factor models are
a good solution, and are commonly used in other industries, such as finance, 
 but one needs to invest in developing factors much
better than the example factors we just mentioned.

\section{Conclusion}

The last section brings us to the point of this writeup.  Having data
from a limited number of channels is a fundamental limit on
information in the Google click-out market.  You can not get around it
by mere calculation.  You need other information sources or aggregation
schemes which may or may not be available.

The points we have touched on are:

\begin{itemize}
\item You can not estimate the variance of individuals from a constant number of aggregated measurements.

This is bad because this interferes with detailed estimates of risk.

\item You can not always undo bad channel assignments by calculation after the fact.

This is bad because this interferes with detailed assignments and 
management of value.
\end{itemize}

In a market information is money.  To the extent you buy or sell in
ignorance you leak money to any counter-parties that know the
things that you do not.  Even if there are no such informed counter-parties 
there are distinct disadvantages in not being able to
un-bundle mixed measurements.  This means it is difficult to un-bundle
mixed sales.  For example we may be making a profit on
a combination purchase of advertisements and we are not able to quickly
determine which advertisements in the combination are profitable and
which are unprofitable.\footnote{By ``quickly determine'' we mean determine
from past data we already have.  What we have shown is we often can not 
determine what we need to know from past data,
but must return to the market with 
new experiments that cost both time and money.}

The capital markets (stocks, bonds, index funds, $\cdots$) 
have evolved and progressed forward
from initial disorganized arrangements to open outcry markets and then to
detailed information environments.  The demands and expectations of
these modern markets include a number of features including:
\begin{itemize}
\item Complete reconciliation and publicly available detailed records of the past.
\item Transparent ``books'' or listings of all current bids and bidders.
\end{itemize}

Not all of these are appropriate for a non-capital market and Google's
on-line advertising markets are just that: Google's.
It is
interesting that before 2007 Yahoo/Overture offered a research interface
that did expose the bidding book.  It will be interesting to see
how the on-line advertising markets evolve and if this feature survives 
in the newer ``more like Google'' Overture market.

The actual lesson we learned in watching others work with on-line
advertising markets are the following.  It is not necessary to be able
to perform any of the calculations mentioned here to run a successful
business.  It is important, however, to have a statistician's
intuition as to what is risky, what can be estimated and what can not
be estimated.  The surprise to the first author that his initial intuition was
wrong, even though he considers himself a mathematician.  It wasn't until
we removed the non-essential details from the problem and found the
appropriate statistical references that we was finally able to fully
convince ourselves that these estimation problems are in fact 
difficult.\footnote{This initial optimism of ours is perhaps a side-effect of a ``can do'' attitude.}

\bibliography{refs}

\section{Appendix}

\subsection{Derivation That a Single Mean is Easy to Estimate}

To show $E[(\hat{a}_w - a_w)^2] =  v_a/n$
we introduce the symbols $x_i$ to denote the 
random variables representing the $n$ apples in our basket and work forward.

To calculate we will need to use some of the theory of
the expectation notation  $E[]$.  Simple facts about the $E[]$
notation are used to reduce complicated expressions into known
quantities.  For example if $x$ is a random variable and
$c$ is a constant than $E[c x] = c E[x]$.
If $y$ is a random variable that is independent of $x$
then $E[x y] = E[x] E[y]$.  And we have for any quantities $x$,$y$
$E[x + y] = E[x] + E[y]$ (even when they are not independent).\footnote{
It is funny in statistics that we spend so much time reminding ourselves that 
$E[x y]$ is not always equal to $E[x] E[y]$ 
that we actually sometimes find it surprising
that $E[x + y] = E[x] + E[y]$ is generally true.
}

Starting our calculation:

\begin{eqnarray}
E[(\hat{a}_w - a_w)^2] & = & 
E\left[\left( (\sum_{i=1}^n x_i)/n - a_w \right)^2 \right]  \label{eq:1} \\
 & = & E\left[\left( \sum_{i=1}^n (x_i - a_w)/n \right)^2 \right] \label{eq:2} \\
 & = & E\left[ \sum_{i=1}^n (x_i - a_w) \sum_{j=1}^n (x_j - a_w) \right]/n^2 \label{eq:3}\\
 & = & E\left[\sum_{i,j} (x_i - a_w) (x_j - a_w)\right]/n^2  \label{eq:4}\\
 & = & E\left[ \sum_{i=1}^n (x_i - a_w)^2 \right]/n^2  \label{eq:5}\\
 & = & E\left[ n (x - a_w)^2 \right]/n^2  \label{eq:6}\\
 & = & E\left[ (x - a_w)^2 \right]/n  \label{eq:7} \\
 & = & v_a/n. \label{eq:8}
\end{eqnarray}

Most of the lines of the derivation are just substitutions or uses of
definition (for example the last substitution on line \ref{eq:8} is
of $E[ (x - a_w)^2 ] \rightarrow v_a$).  A few of the lines use some cute
facts about statistics.  
For example line \ref{eq:4} $\rightarrow$ line \ref{eq:5}
is using the fact that
$E[x_i - a_w] = 0$, which under our independent drawing assumption is 
enough to show $E[(x_i - a_w)(x_j - a_w)] = 0$ when $i \neq j$ 
(hence all these terms can be ignored).  
The line \ref{eq:5} $\rightarrow$ line \ref{eq:6}
substitution uses the fact that each of the $n$ apples was drawn using an
identical process, so we expect the same amount of error in each trial 
(and there are $n$ trials in total).

The conclusion of the 
derivation is that the expected squared error $E[(\hat{a}_w - a_w)^2]$ 
is a factor of $n$ smaller than $v_a = E[(x - a_w)^2]$.  This means 
our estimate $\hat{a}_w$ is getting better and better (closer to the true $a_w$) as we increase the
sample size $n$.

\subsection{Fisher Information and the Cramer-Rao Inequality}

\subsubsection{Discussion}

What is Fisher information?  Is it like the other mathematical
quantities that go by the name of information?

There are a lot of odd quantities related to information each with its own
deep theoretical framework.  For example there are Clausius entropy,
Shannon information and Kolmogorov-Chaiten complexity.  Each of these
has useful applications, precise mathematics and deep meaning.  They
also have somewhat confused and incorrect pseudo-philosophical
popularizations.

Fisher information is not really famous outside of statistics.  
Textbooks motivate it in different ways 
and often introduce an auxiliary function called ``score'' that
quickly makes the calculations work out.  The definition of``score''
uses the fact that 
$\frac{\partial}{\partial \theta} \ln (f(\theta)) = 
\left( \frac{\partial}{\partial \theta} f(\theta) \right) / f(\theta)$
to switch from likelihoods to relative likelihoods.
The entries of the Fisher information matrix are terms of the form
\[
J_{i,j}(\theta) = \int_x f(x;\theta) 
\left( \frac{\partial}{\partial \theta_i} \ln f(x;\theta) \right)
\left( \frac{\partial}{\partial \theta_j} \ln f(x;\theta)  \right)
dx
\]
where $\theta$ is our vector of parameters (set at their unknown true
values that we are trying to estimate) , $x$ ranges over all possible
measurements and $f(x;\theta)$ reads off the likelihood of observing
the measurement $x$ given the parameter $\theta$.

Fisher information is
actually a simpler concept than the other forms of information.
The entries in the Fisher information matrix are merely the expected
values of the effect of each pair of parameters on the relative
likelihood of different observations. In this case, it is showing how alterations in
the unknown parameters would change the relative likelihood of
different observed outcomes.  It is then fairly clever (but not too
surprising) that its inverse can then read off how changes in observed
outcome influence estimates of the unknown parameters.  The Cramer-Rao 
inequality is using Fisher information to describe properties of an inverse
(recovering parameters from observed data) without needing to know the
specific inversion process (how we performed the estimate).

\subsubsection{Calculating Cramer-Rao on the Variance of Variance Estimate}

When attempting to measure the variance of individual apples (Experiment-B)
our data was two sums of random variables (each $x_i$ or $y_i$ representing
a single apple):

\begin{eqnarray*}
T_1 & = & \sum_{i=1}^{n_1} x_i \\
T_2 & = & \sum_{i=1}^{n_2} y_i
\end{eqnarray*}

$n_1,n_2$ can be any positive integers.

Under our assumption that the weight of apples is normally distributed
with mean-weight $a_w$ and variance $v_a$ we can write down the odds-density
for any pair of measurements $T_1,T_2$ as:

\[
f(T_1,T_2;v_a) = 
\frac{1}{2 \pi v_a \sqrt{n_1 n_2}} e^{-(T_1 - n_1 a_w)^2/(2 n_1 v_a) - (T_2 - n_2 a_w)^2/(2 n_2 v_a)}
.\]

To apply the Cramer-Rao inequality we need the Fischer information of this
distribution which is defined as:

\[
J(v_a) = \int_{T_1,T_2} f(T_1,T_2;v_a) 
\left( \frac{\partial}{\partial v_a} \ln f(T_1,T_2;v_a) \right)^2
dT_1 dT_2
.\]

The first step is to use the fact that
\[
\frac{\partial}{\partial x} \ln e^{-f(x)^2} = -2 \frac{\partial}{\partial x} f(x)
\]
and write

\begin{eqnarray*}
J(v_a) & = &
\int_{T_1,T_2} f(T_1,T_2;v_a) 
\left( (T_1 - n_1 a_w)^2/(2 n_1 v_a^2) + (T_2 - n_2 a_w)^2)/(2 n_2 v_a^2) \right)^2
dT_1 dT_2 \\
& = &
\frac{1}{4 v_a^4} \int_{T_1,T_2} f(T_1,T_2;v_a) 
(T_1 - n_1 a_w)^4/n_1^2 
dT_1 dT_2 \\
& & + \frac{1}{4 v_a^4} \int_{T_1,T_2} f(T_1,T_2;v_a) 
2 (T_1 - n_1 a_w)^2 (T_2 - n_2 a_w)^2/(n_1 n_2)
dT_1 dT_2 \\
& & + \frac{1}{4 v_a^4} \int_{T_1,T_2} f(T_1,T_2;v_a) 
(T_2 - n_2 a_w)^4/n_2^2 
dT_1 dT_2 \\
& = & 
\frac{1}{4 v_a^4} \int \Phi_{\sqrt{n_1 v_a}}(x;n_1 a_w) (x- n_1 a_w)^4 dx \\
& &
+ \frac{2}{4 v_a^4} 
\left(\int \Phi_{\sqrt{n_1 v_a}}(x;n_1 a_w) (x- n_1 a_w)^2 dx \right)
\left(\int \Phi_{\sqrt{n_2 v_a}}(x;n_2 a_w) (x- n_2 a_w)^2 dx \right)\\
& & + \frac{1}{4 v_a^4} \int \Phi_{\sqrt{n_2 v_a}}(x;n_2 a_w) (x- n_2 a_w)^4 dx
\end{eqnarray*}
where $\Phi()$ is the standard single variable normal density:
\[
\Phi_\sigma(x;\mu) = \frac{1}{\sqrt{2 \pi} \sigma} e^{-(x-\mu)^2/(2 \sigma^2)}
.\]
The first term is the 4th moment of the normal and it is known that:
\[
\int \Phi_{\sigma}(x;\mu) (x- \mu)^4 dx = 
3 \left( \int \Phi_{\sigma}(x;\mu) (x- \mu)^2 dx \right)^2
.\]
It is also a standard fact about the normal density that 
\[
\int \Phi_{\sigma}(x;\mu) (x- \mu)^2 dx = \sigma^2
.\]
So we have 
\begin{eqnarray*}
J(v_a) & = & \frac{1}{4 v_a^4} \left(
\frac{3 n_1^2 v_a^2}{n_1^2}
+ 2 \left( \frac{n_1 v_a}{n_1} \right) \left( \frac{n_2 v_a}{n_2} \right)
+ \frac{3 n_2^2 v_a^2}{n_2^2}
\right) \\
& = & \frac{2}{v_a^2}
.
\end{eqnarray*}

Finally we have the Fisher Information $J(v_a) = \frac{2}{v_a^2}.$  We
can then apply the Cramer-Rao inequality which says 
that $E[(v_a - \hat{v})^2] \ge 1/J(v_a)$ for {\em any} unbiased estimator
(no matter how we choose $n_1$ and $n_2$)
of $v_a$ (unbiased meaning $E[v_a - \hat{v}] = 0$).  The theory is telling
us that the unknown parameter $v_a$ has such a sloppy contribution to
the likelihood of our observations that it is in fact difficult to
pin down the value from any one set of observations.
In our case we have just shown that 
$E[(v_a - \hat{v})^2] \ge \frac{v_a^2}{2}$, which means no estimation procedure
that uses just a single instance of the total $T_1,T_2$ can reliably 
estimate the variance $v_a$ of individual apple weights.

\subsubsection{Calculating Cramer-Rao Inequality on Multiple Mean Estimates}

In Experiment-C we again have two baskets of fruit- but they contain
apples and oranges in the proportions given by $a_1,o_1,a_2,o_2$.
Our assumption that the individual fruit weights are normally 
distributed with means $a_w,o_w$ and common variance $v$ lets us
us write the joint probability of the total measurements $T_1, T_2$
in terms of the normal-density ($\Phi()$).

For our problem where the variables are the sums $T_1,T_2$ and we 
have two parameters (the two unknown means $a_w,o_w$) and a
single per-fruit variance $v$ we will use the two dimensional 
normal density:
\[
\Phi_{\sqrt{n v}}(T_1,T_2;a_w,o_w) = \frac{1}{2 \pi n v} 
e^{(-(T_1 - a_1 a_w - o_1 o_w)^2 -(T_2 - a_2 a_w - o_2 o_w)^2)/(2 n v)}
.
\]
We concentrate on the variables $T_1,T_2$ and will abbreviate this
density (leaving implicit the important parameters $a_w,o_w,v$) as
$\Phi(T_1,T_2)$.

From this we can read off the difficulty in estimating individual apple
weight:
\begin{eqnarray*}
J_{1,1}(a_w,o_w) & = & 
\int_{T_1,T_2} \Phi(T_1,T_2)
\left( \frac{\partial}{\partial a_w} \ln \Phi(T_1,T_2) \right)
\left( \frac{\partial}{\partial a_w} \ln \Phi(T_1,T_2) \right)
dT_1 dT_2 \\
& = &
\int_{T_1,T_2} \Phi(T_1,T_2)
\frac{
(2 a_1 (T_1 - a_1 a_w - o_1 o_w) + 2 a_2 (T_2 - a_2 a_w - o_2 o_w))^2
}{
4 n^2 v^2
}
dT_1 dT_2 \\
& = &
\frac{a_1^2 + a_2^2}{n v}
\end{eqnarray*}

The first step is using the fact that
\[
\frac{\partial}{\partial x} \ln e^{-f(x)^2} = -2 \frac{\partial}{\partial x} f(x)
\]
The last step is using a number fundamental facts about the normal density:
\begin{eqnarray*}
\int_{x} \Phi_\sigma(x;\mu) dx & = & 1 \\
\int_{x} \Phi_\sigma(x;\mu) (x -\mu) dx & = & 0 \\
\int_{x} \Phi_\sigma(x;\mu) (x -\mu)^2 dx & = & \sigma^2
.
\end{eqnarray*}
These facts allow us say that the
so-called ``cross terms'' (like $(T_1 - a_1 a_w - o_1 o_w) (T_2 - a_2 a_w - o_2 o_w)$)
integrate to zero and the square terms read off the variance. 
One of the reasons to assume a common distribution (such as the normal)
is that almost any complicated calculation involving such distributions 
(differentiating, integrating) can usually be reduced to looking up a few
well know facts about the so-called ``moments'' of the distribution, as
we have done here.  Of, course picking a distribution that accurately models
reality take precedent over picking one that eases calculation.

The other
entries of the Fisher Information matrix can be read off as easily and we
derive:

\[
J(a_w,o_w) = 
\frac{1}{n v}
\begin{bmatrix}
a_1^2 + a_2^2 & a_1 o_1 + a_2 o_2 \\
a_1 o_1 + a_2 o_2 & o_1^2 + o_2^2 \\
\end{bmatrix}
.\]
Substituting our ``typical'' values of $a_1,o_1,a_2,o_2$ from Section \ref{sec:mixture} we have
\[
J(a_w,o_w) = 
\frac{1}{2 v}
\begin{bmatrix}
n  + 4 & n - 4 \\
n - 4 & n + 4
\end{bmatrix}
.\]
At first things look good.  The  $J(a_w,o_w)$ entries are growing with $n$ so 
we might expect the entries of $J^{-1}(a_w,o_w)$ to shrink as $n$ increases.
However, the entries they are all nearly identical 
so the matrix is ill-conditioned and we see larger than expected entries 
in the inverse.  In fact in this case we have:
\[
J^{-1}(a_w,o_w) = 
\frac{v}{8}
\begin{bmatrix}
1 + 4/n & -1 + 4/n \\
-1 + 4/n & 1 + 4/n
\end{bmatrix}
\]
and these entries are not tending to zero- establishing (by the
Cramer-Rao inequality) the difficulty of estimation.

\subsubsection{Cramer-Rao Inequality Holds in General}

By inspecting our last series of 
arguments, we can actually say a bit more.
The difficulty in estimation was not due to our specific
assumed values of $a_1,o_1,a_2,o_2$,
but rather to the fact that the coin-flipping process we 
described earlier
will nearly always land us in about as bad a situation
for large $n$.
We can see that the
larger the differences $|a_1 - o_1|$ and $|a_2 - o_2|$
the better things are for estimation.  The ``strong law of large numbers''
states that as $n$ increases we expect (with probability 1)
to have $|a_1 - o_1| \rightarrow \sqrt{2 v n}$ and
$|a_2 - o_2| \rightarrow \sqrt{2 v n}$.  This means that it would be
very rare (for large $n$) to see differences in $a_1,o_2,a_2,o_2$ larger
than we saw in our ``typical case.''  This lets us conclude that
if there is no constructive bias then
for large $n$ estimation is almost always as difficult as the
example we worked out.

Now if there were any constructive bias 
in the experiment
(such as apples were a bit
more likely in the first basket and oranges were a bit more likely 
in the second basket) then the entries of $J^{-1}()$ would be forced to 
zero and the explicit estimate we gave earlier
would in fact have shrinking error as $n$
grew large.  However only the fraction of the data we can attribute 
to the bias is really helping us (so if it was say a $1/10$th bias
only about $1/10$th of the data is useful to us)  and we would need
a lot of data to experience lowered error (but at least the error
would be falling).  The point is that
the evenly distributed portion of the data is essentially not useful
for inference, and that is why it is so important to be inferring things
that the experiment was designed to measure (and why the limit on 
channel identifiers is bad since it limits the number of things
we can simultaneously design for).

\end{document}